\documentstyle{article}
\begin{document}
\LARGE
\begin{center}
\bf Dispelling the Anthropic Principle from the Dimensionality
Arguments \vspace*{0.7in} \normalsize \large \rm

Zhong Chao Wu

Dept. of Physics

Zhejiang University of Technology

Hangzhou 310032, P.R. China

\vspace*{0.55in}
\large
\bf
Abstract
\end{center}
\vspace*{.1in}
\rm
\normalsize
\vspace*{0.1in}

It is shown that in d=11 supergravity,  under a very reasonable
ansatz, the nearly flat spacetime in which we are living must be
4-dimensional without appealing to the Anthropic Principle. Can we
dispel the Anthropic Principle completely from cosmology?

\vspace*{0.3in}

PACS number(s): 04.65.+c, 11.30.Pb, 04.60.+n, 04.70.Dy


\vspace*{0.5in}

\pagebreak

\rm

\normalsize

Cosmologists would like to use the No-Boundary Universe to predict
everything from First Principles only. On the other hand, the
Anthropic Principle self-imposes the facts in the universe from
the existence of the observer. This is in conflict with the first
principle ambitions of the No-Boundary Proposal. Workers often use
the Anthropic Principle to select dimensionality of spacetime.
However, we show here certain models do not require this.

It is believed that, in M-theory, spacetime has ten or eleven
dimensions, but six or seven dimensions are curved up, leaving
four apparent dimensions we can observe. The orthodox argument for
this phenomenon is based on the Anthropic Principle [1]: there may
exist five or more large dimensions, however only in the
4-dimensional nearly flat spacetime will the question be asked:
``Why is the spacetime 4-dimensional?"

However, under a very reasonable ansatz in $d=11$ supergravity, it
is possible to show that the nearly flat spacetime must be
4-dimensional. Here, in addition to fermion fields, a 3-index
antisymmetric tensor $A_{MNP}$ is introduced into the theory by
supersymmetry [2]. In the classical background of the $WKB$
approximation, one sets the fermion fields to vanish. Then the
action of the bosonic fields can be written
\begin{equation}
\bar{I}= \int \sqrt{-g_{11}}\left (  \frac{1}{2} R - \frac{1}{48}
F_{MNPQ}F^{MNPQ} + \frac{\sqrt{2}}{6\cdot (4!)^2}
\eta^{M_1M_2\cdots
M_{11}}F_{M_1M_2M_3M_4}F_{M_5M_6M_7M_8}A_{M_9M_{10}M_{11}} \right
)d^{11}x,
\end{equation}
where
\begin{equation}
F_{MNPQ} \equiv 4! \partial_{[M}A_{NPQ]},
\end{equation}
\begin{equation}
\eta^{A\cdots N} = \frac{1}{\sqrt{-g_{11}}} \epsilon^{A\cdots N}
\end{equation}
and $R$ is the scalar curvature of the spacetime with metric
signature $(-, +, +, \cdots +)$.

The Kaluza-Klein ansatz we are using is  that spacetime is a
Lorentzian continuation of the product space of two spheres $S_n
\times S_{11-n}$. It is assumed that all components of the $F$
field with mixed indices in the two factor spaces are zero.

The field equations are
\begin{equation}
R_{MN} - \frac{1}{2}Rg_{MN} = \frac{1}{48}
(8F_{MPQR}F_N^{\;\;\;PQR} -g_{MN}F_{SPQR}F^{SPQR}),
\end{equation}
and
\begin{equation}
F^{MNPQ}_{\;\;\;\;\;\;\;\;\;;M}= \left
[\frac{-\sqrt{2}}{2\cdot(4!)^2 }\right ]\cdot \eta^{M_1 \cdots
M_8NPQ}F_{M_1\cdots M_4}F_{M_5\cdots M_8}.
\end{equation}

In the quantum cosmology scenario the Lorentzian evolution of the
universe originates from a compact instanton solution, i.e. a
stationary action solution [3].  Under the ansatz, in the factor
space $S_n \;(n =1,2,3)$ the $F$ components must be vanish due to
the antisymmetry of the indices. Then $F$ must be a harmonic in
$S_{11-n}$ since the right hand side of the field equation (5)
vanishes. It is known in de Rham cohomology that $H^4(S_4) =1$ and
$H^4(S_m) =0 \;\;(m\neq 4)$. So there is no nontrivial instanton
for $n = 1,2,3$. For $n=5,6$, both $F$ components in $S_5$ and
$S_6$ must be harmonics and so vanish since the right hand side of
(5) must be zero. By the dimensional duality, a nontrivial
instanton does not exist for $n= 10, 9, 8$. The case $S_4 \times
S_7$ is the only possibility for the existence of a nontrivial
instanton, the $F$ components must be a harmonic in $S_4$, but do
not have to be in $S_7$. The no-boundary proposal and the ansatz
are very strong, otherwise the nonzero $F$ components could live
in open or closed $n$-dimensional factor spaces $( 4\leq n\leq
10)$, and no explicit restriction is given to the dimensionality
of the universe [4].

Four compact instantons are known [5][6][7][8]. They are products
of a 4-sphere and a round or squashed 7-sphere. These spaces are
distinguished by their symmetries from other infinitely many
solutions with the same $F$ field. From now on, Greek letters run
from 0 to 3 for the indices in $S_4$ and small Latin letters from
4 to 10 for the indices in $S_7$.

One can analytically continue the $S_7$ or $S_4$ space at the
equator to form a 7- or 4-dimensional de Sitter or anti-de Sitter
space, which is identified as our macroscopic spacetime, and the
$S_4$ or $S_7$ space as the internal space. One may naively think,
since in either case the seed instanton is the same, that the
creation of a macroscopic 7- or 4-dimensional universe should be
equally likely. Now it is possible to show (15 years late!) that
the apparent spacetime must be 4-dimensional without appealing to
the Anthropic Principle [9].

The solutions of the $F$ components in the $S_4$ factor space must
be
\begin{equation}
F_{\mu \nu \sigma \delta} = i\kappa \sqrt{g_4}\epsilon_{\mu \nu
\sigma \delta },
\end{equation}
where $g_4$ is the determinant of the $S_4$ metric and $\kappa$ is
a constant. If the other $F$ components are zero, then we obtain
the Freund-Rubin model [5]. There the $F$ field plays the role of
an anisotropic effective cosmological constant, which is
$\Lambda_7 = \kappa^2/3$ for $S_7$ and $\Lambda_4 = - 2\kappa^2/3$
for $S_4$, in the sense that $R_{mn} = \Lambda_7 \; g_{mn}$  and
$R_{\mu \nu} = \Lambda_4 \; g_{\mu \nu}$, respectively. The $S_4$
space must have radius $r_4 = (3/\Lambda_4)^{1/2}$ and metric
signature $(-,-,-,-)$, while the $S_7$ space is of radius $r_7
=(6/\Lambda_7)^{1/2}$ and metric signature $(+,+, \cdots +)$.

The $S_4$ metric can be written
\begin{equation}
ds_4^2= -dt^2  - \frac{3}{\Lambda_4} \sin^2\left
(\sqrt{\frac{\Lambda_4}{3}}t \right )(d\chi^2 + \sin^2 \chi
(d\theta^2 + \sin^2 \theta d\phi^2)).
\end{equation}
One can obtain the 4-dimensional anti-de Sitter space through two
steps. First, one has to analytically continue on a three surface where the
metric is stationary. One can choose $\chi = \frac{\pi}{2}$ as the
surface, set $\omega = i(\chi - \frac{\pi}{2})$, and then
analytically continue the metric through the null surface at $t =
0$ by redefining $ \rho = \omega +\frac{i\pi}{2}$ and get the
anti-de Sitter metric
\begin{equation}
ds_4^2= -dt^2  + \frac{3}{\Lambda_4} \sin^2\left
(\sqrt{\frac{\Lambda_4}{3}}t \right )(d\rho^2 + \mbox{sinh}^2 \rho
(d\theta^2 + \sin^2 \theta d\phi^2)).
\end{equation}

The relative probability of the creation, at the $WKB$ level, is
the exponential to the negative of the Euclidean action of the
instanton $S_7 \times S_4$
\begin{equation}
P =\Psi^* \cdot \Psi \approx \exp (-I) ,
\end{equation}
where $\Psi$ is the wave function of the configuration at the
quantum transition. The configuration is the metric and the
matter field at the equator. $I$ is the Euclidean action.

If we are living in the Lorentzian section of the 7-dimensional
(4-dimensional)  factor space with $S_4 (S_7)$ as the internal
space, then the Euclidean action $I$ should take the form
\begin{equation}
I=- k\int \sqrt{g_{11}}\left (  \frac{1}{2} R - \frac{1}{48}
F_{MNPQ}F^{MNPQ} + \frac{\sqrt{2}i}{6\cdot (4!)^2}
\eta^{M_1M_2\cdots
M_{11}}F_{M_1M_2M_3M_4}F_{M_5M_6M_7M_8}A_{M_9M_{10}M_{11}} \right
)d^{11}x,
\end{equation}
where $k = 1 (-1)$. This is obtained through analytical
continuation from the Lorentzian section as in the usual
4-dimensional Euclidean quantum gravity. The choice of the value
$k$ is also supported by its cosmological implication. The $-
\frac{k}{2}R$ term in the action can be decomposed into the
difference of the curvatures for the two factor spaces $-
\frac{k}{2}(R_7 - R_4)$. The negative sign in front of $R_7 (R_4)$
is required so that the perturbation modes of the gravitational
field in the $S_7$ background would take the minimum excitation
state allowed by the Heisenberg Uncertainty Principle.

The Euclidean action $I$ of  the $AdS_4 \times S_7$ space can be
calculated [9]
\begin{equation}
I =\frac{1}{3}\kappa^2 V_7V_4 ,
\end{equation}
where $V_7$ $\;(V_4)$ is the volume of $S_7$ $\;(S_4)$.

The field equation (5) is derived from the action (1) for the
condition that the tensor $A_{MNP}$ is given at the boundary.
Therefore, if one uses the action (1) in the evaluation of the
wave function and the probability, then the induced metric and
tensor $A$ on it must be the configuration of the wave function.
The wave function is expressed by a path integral over all
histories with the configuration as the only boundary. In deriving
Eq. (9), one adjoins the histories in the summation of the wave
function to their time reversals at the equator to form a manifold
without boundary and discontinuity.

The induced metric and scalar field (if there is any) at the
equator will remain intact under the time reversal operation.
However, for other fields, one has to be cautious. This occurs to
our $A_{MNP}$ field. If one use $A_{MNP}$ as the argument for the
wave function, then one has implicitly fixed the gauge condition
and there must be a discontinuity at the equator $\chi
=\frac{\pi}{2}$ which cannot be fixed by a gauge transformation.

In order for the instanton approach to be valid, one has to use
the canonical conjugate representation. One can make a Fourier
transform of the wave function $\Psi (h_{ij}, A_{123})$ to get the
wave function $\Psi (h_{ij}, P^{123})$,
\begin{equation}
\Psi ((h_{ij}, P^{123}) = \frac{1}{2\pi} \int_{-\infty}^{\infty}
e^{iA_{123}P^{123}} \Psi(h_{ij}, A_{123}).
\end{equation}
where $P^{123}$ is the canonical momentum conjugate to $A_{123}$.
the only degree of freedom of the matter content under the
minisuperspace ansatz is $A_{123}$ for a specified gauge.

The discontinuity which occurs at the instanton equator is thus
avoided using the momentum representation. At the $WKB$ level, the
Fourier transform of the wave function is equivalent to the
Legendre transform of the action. The Legendre transform has
introduced an extra contribution  $-2 A_{123}P^{123}$ to the
Euclidean action, where all quantities are in the Euclidean
version, and the factor 2 is due to the two sides of the equator
in the adjoining. Then the effective action becomes [9]
\begin{equation}
I_{effect} = - \frac{2}{3} \kappa^2 V_7V_4.
\end{equation}

If we consider the quantum transition to occur at the equator of
$S_7$ instead, using the same argument, then it turns out that the
corresponding canonical momentum using the time coordinate in
$S_7$ vanishes, and the effective action should be the negative of
(11).

Since the creation probability is the exponential to the negative
of the Euclidean action, the probability of creating a
7-dimensional macroscopic universe is exponentially suppressed
relative to that of the 4-dimensional case.

In the classical framework, the $S_7$ factor space in the
Freund-Rubin model can be replaced by $S_2\times S_5$, $S_2 \times
S_2 \times S_3$, $S_4 \times S_3$ or other Einstein spaces.
However, all these product spaces have volumes smaller than that
of $S_7$. This would lead to an exponential suppression of the
creation probability. Therefore, the internal space must be the
round $S_7$ space.

In the Freund-Rubin model, the $S_7$ factor space can be replaced
by a general Einstein space with the same cosmological constant
$\Lambda_7$. Among them, the Awada-Duff-Pope model [7] is most
interesting, where the round 7-sphere is replaced by a squashed
one. As far as the scenario of quantum creation is concerned , the
argument for the Freund-Rubin model remains intact; the only
alternations are that the quantum transition should occur at one
of its stationary equators and $V_7$ should be the volume of the
squashed 7-sphere.

The same argument applies to the Englert model [6]. There, in
addition to the $F$ components of the space $S_4$ in (6), the
$F_{mnpq}$ components of the $S_7$ space are non-vanishing. As in
the Freund-Rubin model, before we take account of the Legendre
term, the Euclidean action of the Englert $AdS_4 \times S_7$ space
is [9]
\begin{equation}
I =-\frac{1}{4} \kappa^2V_7V_4.
\end{equation}
After including the Legendre term the effective action becomes [9]
\begin{equation}
I_{effect} =-\frac{2}{3} \kappa^2V_7V_4.
\end{equation}

If the quantum transition occurred at an equator of the $S_7$
space, one has to include the Legendre terms correspondingly. In
contrast to the Freund-Rubin model, the canonical momenta do not
vanish. Fortunately, due to the symmetries of the momenta, the sum
of the $C^3_6 = 20$  Legendre terms cancel exactly. The action is
the negative of that in (14). Again, one can conclude that the
universe we are living is most likely 4-dimensional.

Englert, Rooman and Spindel also discussed the model with a
squashed $S_7$ factor space [8]. It is believed that our
conclusion should remain the same.

The right configuration for the wave function  has also been
chosen in the problem  of quantum creation of magnetic and
electric black holes [10].  If one considers the quantum creation
of a general charged and rotating black hole, this point is even
more critical. It is become so acute that unless the right
configuration is used, one cannot even find a constrained
instanton seed [11].

For $d=11$ supergravity, there is no way to discriminate the $d=4$
and $d= 7$ macroscopic universes in the classical framework, as in
other similar but more artificial models. This discrimination can
be realized only through quantum cosmology.

Now, it seems that one can dispel the Anthropic Principle so far
as the dimensionality of the spacetime is concerned. Can we show
the Anthropic Principle is not required at all?

\vspace*{0.1in}

\bf References:

\vspace*{0.1in}
\rm

1. Hawking S W 2001 \it The Universe in a Nutshell  \rm (Bantam
Books, New York)chap 3

2. Cremmer E, Julia B and Scherk J 1978 \it Phys. Lett. \bf
B\rm\underline{76} 409

3. Hartle J B and Hawking S W 1983 \it Phys. Rev. \rm \bf D\rm
\underline{28} 2960

4. Wu Z C 1984 \it Phys. Lett. \bf B\rm\underline{146} 307 (1984).
Hu X M and Wu Z C 1984 \it Phys. Lett. \bf B\rm\underline{149} 87
; 1985 \it Phys. Lett. \bf B\rm\underline{155} 237 ; 1986 \it
Phys. Lett. \bf B\rm\underline{182} 305

5. Freund G O and Rubin M A 1980 \it Phys. Lett. \bf
B\rm\underline{97} 233

6. Englert F 1982 \it Phys. Lett. \bf B\rm\underline{119}, 339

7. Awada M A, Duff M J and  Pope C N 1983 \it Phys. Rev. Lett.
\rm\underline{50} 294

8. Englert F, Rooman M and Spindel P 1983 \it Phys. Lett. \bf
B\rm\underline{127} 47

9. Wu Z C 2002 \it Gene. Rel. Grav.  \rm \underline{34} 1121,
hep-th/0105021

10. Hawking S W and Ross S F 1995 \it Phys. Rev. \rm \bf
D\rm\underline{52} 5865. Mann R B and Ross S F 1995, \it Phys.
Rev. \rm \bf D\rm\underline{52} 2254

11. Wu Z C 1997 \it Int. J. Mod. Phys. \rm \bf D\rm \underline{6}
199, gr-qc/9801020; 1999 \it Phys. Lett. \bf B\rm \underline{445}
174, gr-qc/9810012

\end{document}